# Post-collision hydrodynamics of droplets on cylindrical bodies of variant convexity and wettability

**Gargi Khurana, Nilamani Sahoo** and **Purbarun Dhar**\*

Department of Mechanical Engineering, Indian Institute of Technology Ropar,

Rupnagar–140001, India

\* Corresponding author:
E–mail: purbarun@iitrpr.ac.in ; pdhar1990@gmail.com
Phone: +91-1881-24-2173

## Abstract

Droplet impact, dynamics, wetting and spreading behavior on surfaces imposes rich and interesting physics, in addition to extensive understanding of processed employing droplets and sprays. The physics and mechnisms become richer and more interesting in the event the geometry, morphology and wettability of the surface provides additional constraints to the fluid dynamics. Post-impingement morphology and dynamics of water droplets on convex cylindrical surfaces of variant radii has been explored experimentally. Droplet impact and post-impact feature studies have been conducted on hydrophillic and superhydrophobic (SH) cylindrical surfaces. Effects of the impact Weber number (We) and target-to-drop diameter ratio on the wetting and spreading hydrodynamics has been studied and discussed. The post-impact hydrodynamics have been quantified  employing dedicated non-dimensional variables, such as the wetting fraction, the spreading factor and non-dimensional film thickness at the north pole of the target. The observations reveal that the wetting fraction and spread factor increases with an increase in the impact We  and decrease in the target-to-drop diameter ratio. An opposite trend is noted for the non-dimensional film thickness at the target's north pole. It is also deduced that the spread factor is independent of the target wettability whereas the wetting fraction is remarkably low for SH targets. The lamella dynamics post spreading has also been observed to be a strong function of the wettability, impact We and the diameter ratio and the same has been explained based on wetting and inertial principles. An analytical expression for temporal evolution of film thickness at north pole of the cylinderical target is derieved from first priciples. The article also proposes a theoretical model based on energy conservation for predicting the maximum wetting fraction for variant cylindrical targets in terms of the governing Weber number (We) and Capillary number (Ca). It is observed that the experimental measurements are in good agreement with the theoretical predictions.

## 1. Introduction

The hydrodynamics of droplet impact on solid surfaces has been studied extensively by many researchers for its scientific interest and intrinsic physical beauty. Understanding droplet post-impact dynamics is relevant to various industrial and environmental processes as well. Fluidized catalytic cracking, inkjet printing (van Dam and Le Clerc 2004; Haskal et al. 2002), surface coating, spray quenching, pesticide spraying, fuel injection (Ueda et al. 1979; Moreira et al. 2010), fire suppression sprinklers and spray scrubbing of particulate matters are a few applications where droplet and surface interaction is an important governing factor. The field of droplet impact hydrodynamics is a fluid dynamics phenomenon of a single droplet striking a surface and its consequent behavior. The outcome of this hydrodynamic interaction depends complexly on the combined influence of the droplet size, liquid properties, surface properties, surface geometry and impact conditions .(Crooks et al. 2001; Park et al, 2003; Roux and Cooper-White 2004). Spreading and recoiling dynamics has been observed and modelled by various researchers in an attempt to interpret the collisional deformation of the droplet on a surface and its subsequent wetting dynamics. Droplet impact on flat surfaces has been an extensively investigated area of research with various comprehensive reviews (Yarin 2006; Rein 1993; Josserand and Thoroddsen 2016) available on the subject matter. These comprehensive reviews discuss the hydrodynamics, wetting regimes and spreading phenomena on flat surfaces of variant wettability or surface morphologies, such as roughness or texture. Theoretical models for predicting the maximum spreading diameter based on energy balance considerations have been proposed in several studies (Chandra and Avedisian 1991; Pasandideh-Fard et al. 1996; Mau et al. 1997; Mitra et al. 2016) as it is among the major criteria affecting the role of such droplets in targeted applications as discussed earlier.

It is only in recent times that studies have evolved towards understanding the behavior of droplet impact onto surfaces which are not planar, since such surfaces are more likely to be encountered in manufacturing and precision utilities, such as spray coating, spray polishing, inkjet coloring, etc. compared to flat surfaces. Recent studies have investigated droplet impact on to spherical target surfaces. Hardalupas et al. (1999) reported an experimental study of liquid (ethanol and glycerol solutions in water) droplet impact on small solid spheres having diameter ratios (target to droplet diameter) in the range of 0.12 to 0.29 and high impact velocities. The study reported formation of finger like or inverted crown shaped structures upon impact, which subsequently disrupt due to capillarity. The study also concluded that the onset of splashing is favored by increase in surface curvature. Bakshi et al. (2007) reported the spatial and temporal variation of the film thickness on spherical target surfaces for liquid (water and isopropanol) droplet impact at low velocities. The diameter ratios considered in the study ranged from 0.15 to 0.81. The experimental results indicated three temporal phases of the film dynamics, namely, the initial drop deformation phase; the inertia dominated spreading phase, and the viscosity dominated phase. The effect of impact Reynolds number (Re) and diameter ratio was studied and it was found that in the first two stages, the non-dimensional temporal variation of film thickness for different values of Re collapses onto a single master curve, indicating independent behavior from the Re at lower impact velocity regimes. Additionally, the transition to the third phase is observed to occur earlier for lower Re conditions. An analytical expression for film thickness has also been proposed for the inertia dominated phase. Mitra et al. (2013) reported a theoretical and experimental study of subcooled droplet impacting on spherical brass targets in the temperature range of 20 to 250 $^o$C. Droplet spread factor was investigated over a range of Weber numbers (We) and the maximum spread was found to be in agreement with the Chandra & Avedisian (1991) model. The study reported wetting contact for surfaces at room temperature whereas at elevated temperature regimes, wetting is arrested due to the formation of a thin vapor cushion at the interface of surface and liquid (the Leidenfrost effect). Additionally, the droplet contact time was found to decrease with increase in the impact We.

Banitabaei et al. (2017) studied the effect of impact velocity and wettability of the particle on the morphology of droplet impacting on to small spherical particles (diameter ratio more than 1). The study considered various post-impact geometrical parameters, viz. the film thickness, lamella height and lamella base diameter. It was found that the lamella formation is only possible when a droplet impacts on a hydrophobic particle of some appropriate diameter ratio and impact velocity. Hung & Yao (1999) experimentally investigated impact of water droplets of a single size on cylindrical wires. Disintegration and dripping were reported as the outcome by parametrically varying wire size and impact velocity. It was found that higher impact velocity and smaller wire size favors disintegration mode. Dripping was further classified into momentum induced and gravity induced modes. Sher et al. (2013) studied the factors that affect the amount of liquid trapped upon a dry horizontal wire. A non-dimensional criterion for critical eccentricity value at which mass of liquid trapped is maximum was reported. Jin et al. (2017) noted that the radius of the cylindrical surface had more influence on the maximum spreading diameter in the azimuthal direction than in the axial direction. Liu et al. (2017) adopted the coupled level set and volume-of-fluid (VOF) method to simulate impact process on tubular geometries for different hydrophobicity and impact velocities. It was found that with increasing contact angle, the spreading diameter decreases while the height of the liquid film at the center increases.

Hydrodynamics of droplets upon impact has also been simulated using the lattice Boltzmann method and four stages, namely, moving, spreading, nucleating, and dripping were reported (Shen et al, 2012) for curved surfaces. Collisional dynamics of liquid droplets on curved geometries is a neglected area of research with significant possibilities to explore numerous outcomes, which predominantly depend on the droplet-target size ratio, impact conditions, wettability, etc. Compared to planar cases, droplet impact on curved surfaces is not well-studied and the understanding of the hydrodynamics is scarce. Droplet dynamics on a curved surface is important in applications such as impact of water drops on aircraft wings and has implications towards understanding consequent ice-formation dynamics, spraying and coating of complex geometry objects, coating of turbine blades, automotive parts, etc. Additionally, the hydrodynamics on curved features would be rich in its physics. The present article experimentally explores the post-impact hydrodynamics of water droplets on cylindrical targets of variant diameter ratios and wettability. The spreading dynamics, wetting behavior, film drainage behavior at the north pole of the cylinder, and the lamellae dynamics beyond the south-pole have been discussed in depth. Imaging from both the side and top reveals several interesting dynamics that the droplet undergoes during the post-impact phase. Further, an analytical model has been proposed to quantify the film drainage behavior with time as well as determine the effective spread of the droplet under various impact conditions, and good agreement with experimental observations has been achieved.

## 2. Materials and methodologies

A custom arranged experimental setup has been used in the present studies and Fig. 1 illustrates the schematic of the experimental setup used. A 250µl chromatography syringe (with a stainless steel 22G gauge needle), attached to a digitally controlled, precision drop dispenser mechanism (Holmarc Opto-Mechantronics Pvt. Ltd., India) is employed to generate water droplet which falls freely from a desired height onto the cylindrical target (polished stainless steel rods of various diameters). A high speed camera (Photron, UK) mounted with a G-type AF-S macro lens of constant focal length 105mm (Nikkor, Nikon) is used to capture the impact phenomena at 3600 frames per second (at 1024 x 1024 pixels resolution). Every impact case is has been conducted for two camera settings, to obtain the dynamics from both top and front views. In the case of front view, the camera is placed coaxially with the cylindrical target as shown in the fig. 1. For the top view, the camera is placed vertically and orthogonally to the axis of the target surface. A brightness controlled white LED (light emitting diode) backlight (Holmarc Opto-mechatronics, India) is used for illumination and is placed such that camera, the target and the centre of the backlight are in a straight line.

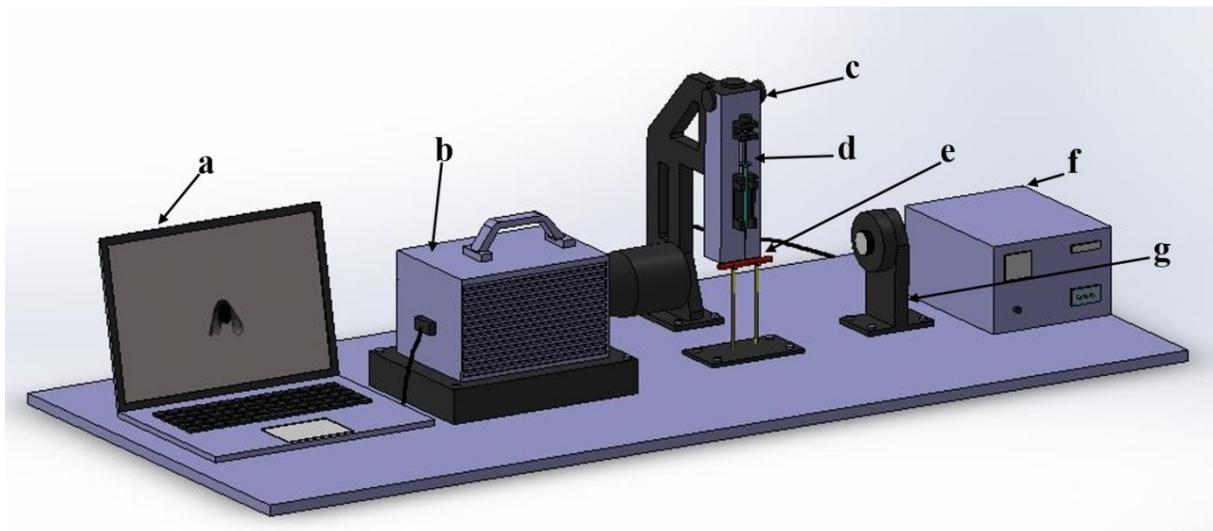

**Fig. 1:** Schematic of the experimental setup (a) computer for data acquisition and control of the camera unit (b) high speed camera with a 105 mm macro lens assembly (for top view experiments, the camera is mounted vertically using a tripod) (c) precision microliter droplet dispensing mechanism (d) chromatography syringe with stainless steel needle (e) cylindrical target body (f) droplet dispenser unit controller (g) LED (light emitting diode) illumination unit with intensity controller.

The diameter of the water droplet before impact is maintained as ~ 2.9 ± 0.2 mm. It is determined by dispensing a fixed volume of water by the digitized dispenser and also verified by weighing the dispensed droplet using a precision electronic balance (Shimadzu ATX, Japan). The height of the tip of the needle from the cylindrical target body is varied to provide different droplet impact velocities of ~0.95m/s, ~1.17m/s and ~1.5m/s (all values are accurate to within ± 5 %). The velocities have been determined from image processing of the droplet during its free fall moments before impact. It was ensured that droplet impacts exactly at the centreline of the cylindrical target by positioning the target body accurately employing trial and error method. The target surfaces used for the experiments are stainless steel rods of diameters 1.54mm, 2.4mm and 4mm (diameters are ensured using digital Vernier callipers), which are cleaned with acetone and then dried in hot air oven. Another set of similar rods is spray-coated with a commercial superhydrophobic (SH) concoction (Ultra Tech International Inc., USA) to produce SH rods. The coating thickness ranges in few microns, and hence does not affect the diameter of the rods. The images of the collisional events are post processed using the open source software ImageJ software to quantify various parameters, viz. wetting fraction, spread factor and film thickness at the north pole of the target surface. Illustrations of these parameters have been provided in and defined in fig 2. In order to ensure the repeatability of the experiments, each impact was repeated thrice on three randomly selected regions on the target bodies. All experiments were performed at temperature $27 \pm 5^0 C$, and relative humidity of $55 \pm 5\%$. Additionally, typical values of the static contact angle for the hydrophilic and SH targets have been tabulated in Table 1 along with the values of surface energies (the liquid-gas and solid-liquid components).

**Table 1:** The liquid-gas and solid-liquid components of surface energy and static contact angle for cylindrical targets of variant wettability)

| Cylindrical target | $\sigma_{sl}$ (J/m$^2$) | $\sigma_{lg}$ (J/m$^2$) | Contact angle |
|---|---|---|---|
| Stainless steel | 0.18[a] | 0.07286 | $45 \pm 3^o$ |
| SHS | 0.2726 | 0.07286 | $135 \pm 2^o$ |

[a] Omar et al. (2015)

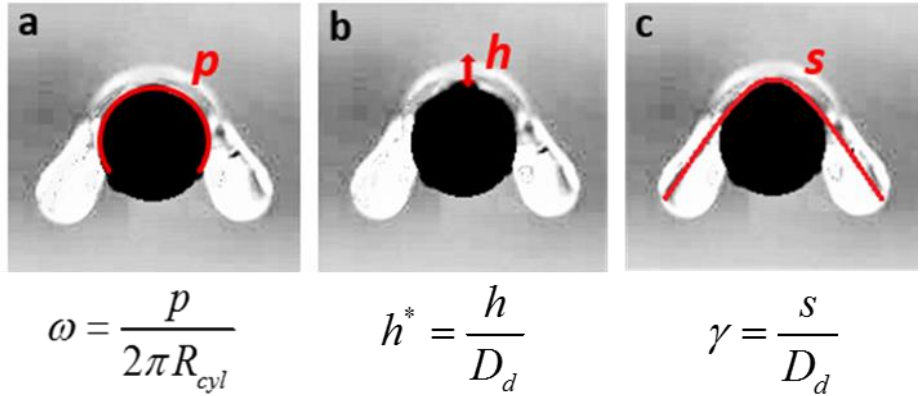

**Fig. 2:** Illustration of the transient (a) wetting fraction, $\omega$ (b) non-dimensional liquid film thickness, $h^*$ and (c) spreading factor, $\gamma$. $P$ denotes the wetted perimeter; $R_{cyl}$ denotes the radius of the target and $D_d$ denotes the diameter of the droplet before impact.

## 3. Results and discussions

The experimental results have been presented in the form of temporal variations of three non-dimensional parameters, viz. the wetting fraction, the non-dimensional film thickness at the north pole of the target and the spread factor, which are illustrated in fig. 2. Measurements of p, h and s are done via image processing using the open source processor ImageJ. In the present study, experiments have been done for two wettability and three impact velocities and three diameter ratios. The diameter ratio $D^*$ is defined as $D^* = \frac{2R_{cyl}}{D_d}$ where $R_{cyl}$ is radius of cylindrical target. For all cases, the impact Weber number is defined as $We = \frac{\rho v^2 D_d}{\sigma}$, where $\rho$ is the density of the liquid, $v$ is the velocity at impact (determined from image processing), $D_d$ is the diameter of the droplet at impact and $\sigma$ is the surface tension of the liquid with respect to air. Throughout the study, the We has been changed by varying the impact velocity of the droplet and $D^*$ is changed by varying the values of $R_{cyl}$.

### 3.1. Hydrophilic surfaces

The post-impact dynamics have been qualitatively presented in the form of time series arrays. The arrays illustrated in Figs. 3 and 4 show the deformation dynamics, wetting behaviour and post-wetting lamella dynamics of a water droplet on a cylindrical surface. Fig. 3 illustrates a sequence of the front view images at different times for water droplet impact on hydrophilic cylindrical targets (diameter 2.4mm) with different impact We. It is observed that with the wire diameter remaining constant, impact at higher velocities causes the wire to be completely engulfed by the droplet, whereas at lower velocities, the engulfing is only partial. While the target diameter and the wettability remain unchanged, the inertia of the impact governs the hydrodynamics. At lower We, (We = 36), the inertia of the droplet post impact is low and the spreading ceases within 5 ms, and the next stage of film drainage is evident from the array. Beyond 5 ms, the extent of spread remains same and the drainage of the film leads to formation of the lamella below the cylinder. It is also noteworthy that the liquid film does not meet and coalesce at the south-pole, and the two lamellae remain distinctly separated. On the contrary, the spreading is evident to continue in cases of higher We till the point the lamellae coalesce and the whole surface can be seen to be wetted (11.12 ms at We = 54 and 6.95 ms at We = 89). It is also evident that the regime of spreading and the governing We determines the nature and behaviour of the lamella post spreading. At high We, the lamellae on both ends coalesce due to the large inertial spreading regime and the single column of liquid exhibits capillary instability (We = 89, 9.17–15.01 ms). The column eventually breaks up into smaller constituent daughter droplets, characterized by typical necking (13.07 and 15.01 ms) which is similar in behaviour to the typical Plateau-Rayleigh instability of a liquid jet.

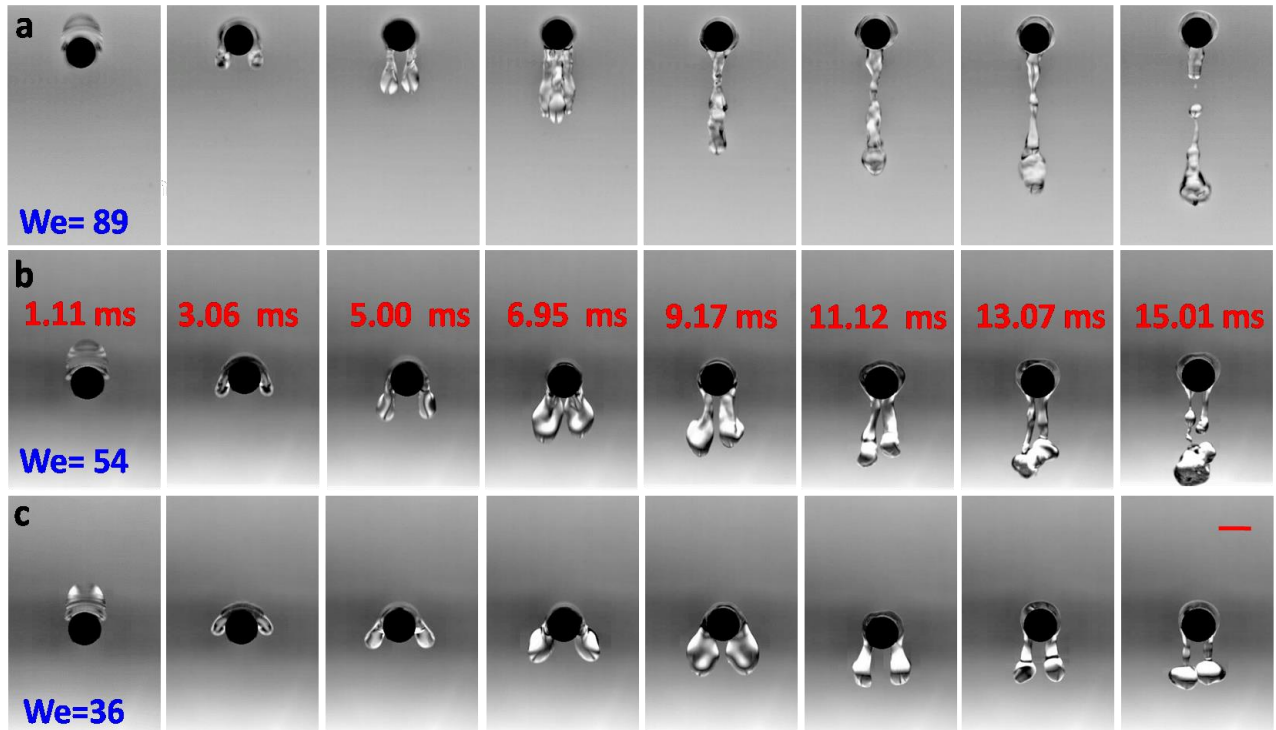

**Fig. 3:** Post-impact deformation (front view), wetting and post-wetting lamella dynamics water droplet on hydrophilic cylindrical target surface (diameter 2.4mm) at impact velocities of (a) 1.5m/s, (b)1.17m/s and (c) 0.95m/s. The associated We and the time frames have been illustrated. The magnitude of scale bar (bottom right) is 2.4mm. The array illustrates the role of the impact We.

The noteworthy behaviour comes to the forefront in case of the moderate We impact. While the inertial spreading leads to coalesce of the film (We = 54, 11.12 ms), the lamellae do not coalesce. Instead, two distinct liquid columns appear, which fuses only at the end of the bulging phase, and detach as a single liquid mass due to capillary instability (13.07–15.01 ms). At even lower impact We, the inertial spreading is low and the film does not wet the whole periphery of the target. This causes the lamellae to evolve without wetting and coalescence disturbances, leading to the formation of symmetric bee-wing lamellae (We = 36, 9.17 ms). Since the periphery is not fully wetted, the lamellae remain well separated and detach off as independently (at the same time) without coalescence. The absence of lamella coalescence also ensures that the capillary instability before the detachment event is largely reduced in strength, and the two lamellae detach in the form of two minor droplets (15. 01 ms). Another important physical mechanism which is noteworthy is the subsequent film replenishment due to wetting recoil. It can be observed in each case (We = 89, 11.12 ms; We = 54, 9.17 ms and We = 36, 9.17 ms) that the film of liquid near the north pole of the target depletes down to a minimum thickness, and then regains a thicker morphology. This is analogous to the phase at which the lamella development terminates and the lamella detachment phase initiates. During the lamella development regime, the weight of the growing lamella drains the film to a minimum. Beyond this point, the wetting tension between the film and the hydrophilic surface prevents further film drainage. At this juncture, the lamellae are forces to begin its detachment process due to capillary instability, caused by the interactions between the wetting tension at the target and the weight of the lamella. Once the major portion of the lamella begins to detach by necking or threading, the remaining mass of fluid shrouding the target experiences recoil due to capillarity at the neck, leading to partial replenishment of the film near the north pole. It is observable from all the cases that post-detachment of the lamellae leads to increment in thickness of the drained film.

The lamella formation and dynamics is observed to be a strong function of the impact conditions and the post impact structures are different in each case. Fig 4 illustrates the hydrodynamics of the droplet on cylindrical targets of different diameter ratios (D*) for evolving time. It is observed that increase in the D* leads to drastic arrest of the post-impact lamella formation and its hydrodynamics. At low values of D*, the droplet evolves into a full-fledged butterfly wing shaped lamella and the lamellae further coalesces before the droplet detaches off the southern region of the target, similar to a dripping fashion by forming a chain of microscale droplets. Increase in the D* arrests the coalescence, and bulbous or ear-ring shaped lamella are generated before detachment. Additionally, the dripping behaviour post lamellae detachment is also arrested. At highest D*, it can be observed that proto-lamella formation occurs, but the surface hydrophilicity wettability and target size arrests evolution of the spreading and lamella formation, causing the droplet to split up and form three static daughter droplets arranged around the periphery of the target.

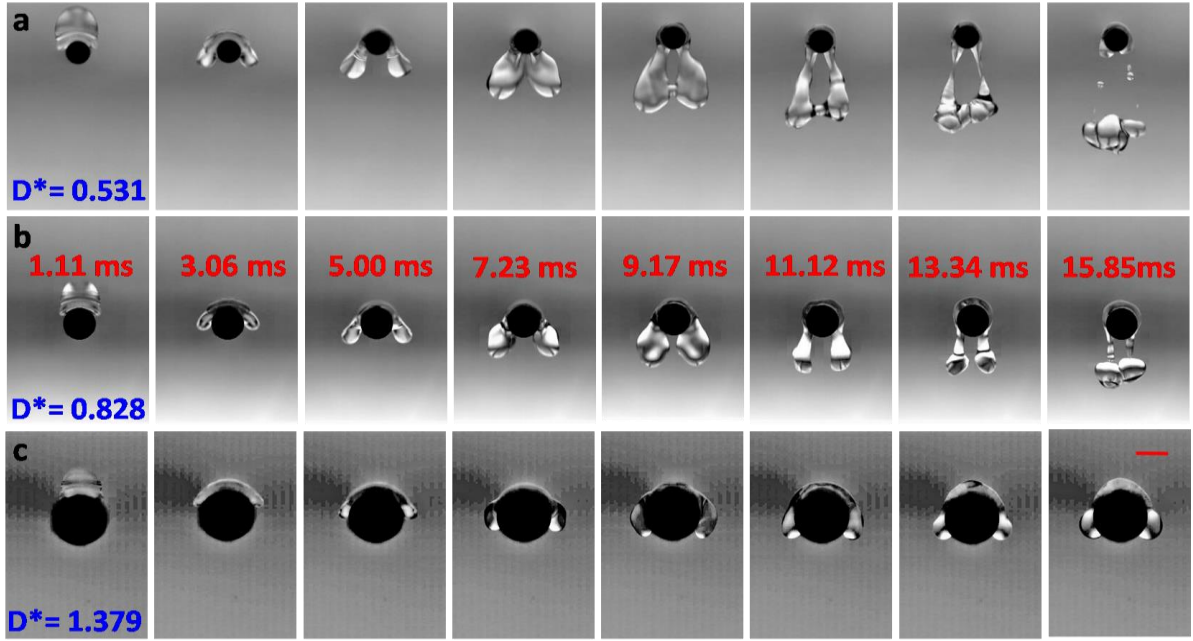

**Fig. 4:** Front view of post-impact images of droplet onto hydrophilic cylindrical target surfaces of diameters a) 1.54mm, b) 2.4mm and c) 4 mm at 0.95m/s impact velocity (impact We = 36). The magnitude of scale bar is equal to 2.4mm (in bottom right image). The array illustrates the role of D*.

Fig. 5 illustrates the temporal variation of wetting fraction on cylindrical targets of different sizes. The wetting fraction $\omega$ is expressed as the ratio of wetted perimeter of the cross section of the cylindrical target to the total perimeter of the cross section of the (illustrated in fig 2.a.). $\tau$ is the non-dimensional time, represented as $\tau = \dfrac{tV}{D_d}$, where $t$ is the time evolution from the instant of impact and $V$ is the impact velocity. The instant $\tau = 0$ corresponds to the moment the droplet just comes in contact with the target surface. In fig 5A, the wetting fraction for different $D^*$ has been illustrated for impact We=36. The wetting fraction is observed to increase up to a certain time period to attain a maximum value and thereafter decreases during the recoiling phase. Considering the highest value of D*, it is observed that the wetting fraction attains a plateau, indicating that the droplet spreads and obtains a stable and stationary configuration on the target (as seen in inset c). As the D* reduces, the wetting fraction is throughout higher over the whole time frame. Furthermore, the point of maximum wetting is also attained faster for lower values of $D^*$ since at a fixed We, the engulfment process is faster at lower D*. At lower D*, the wetting process terminates into the lamellae formation and subsequent lamella detachment due to capillary instabilities. As discussed earlier, lamella detachment is associated with capillary recoil, which suddenly pushes back the film of fluid away from the south pole of the droplet by some extent. This behaviour leads to a partial depletion of the wetted film near the south pole and has been described by a reduction in the effective wetting fraction in fig. 5 A. Since the capillary recoil is stronger for lower D* (due to more vigorous capillary instabilities during the lamella detachment), the recoil observed after the wetting maxima is highest for the lowest D*.

Fig 5 B illustrates a plot similar to 5A, but for impact We=89. In accordance to the above discussion, the wetting fraction is initially higher for lower $D^*$, but unlike for lower We, the maximum wetting fraction at high We for all $D^*$ eventually becomes 1 and no visible recoil from the maximum is noticeable. This is caused by the higher inertia of impact, which causes the lamellae to coalesce due to inertial spreading. Unlike the capillary coalescence in the lower We cases, the inertial spreading induced coalescence leads to vigorous inertia-capillary instabilities in the fused lamellae (inset a and b of 5 B). The images corresponding to maximum wetting fraction for each case have been shown as insets in the respective plots. The largest target diameter, as observed in inset c 5A, essentially prevents the drop from spreading towards the south-pole, as the larger perimeter is capable of arresting the wetting due to surface shear. However, the same target, in the event of higher We, is unable to arrest the spreading due to the large inertial spreading regime at higher We. The phases of spreading can be essentially deduced from the time plots based on the effective slope of the wetting curve. It can be observed that the curves contain an initial small region of sharper slope, followed by a region of flattened slope, followed again by a minute region of sharp increasing slope before the peak wetting is reached. The initial region of sharp slope represents the typical inertial phase of spreading, and it can be observed that for a particular D*, this region is steeper for higher We cases than lower We cases, since higher We cases impose faster inertial spreading.

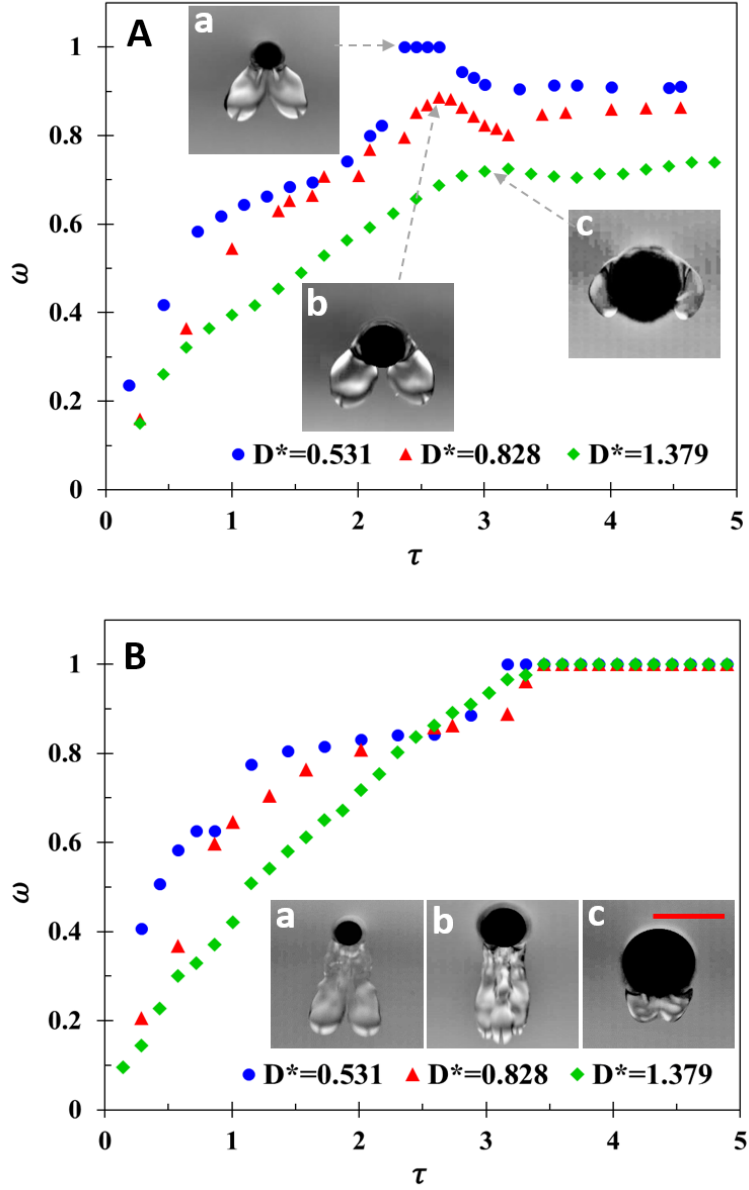

**Fig.5**: Temporal variation of wetting fraction on hydrophilic cylindrical target surfaces of different D* (A) We= 36. Inset: Front view post impact images depicting maximum wetting fraction for (a) D*=0.531, τ=2.37 (b) D*=0.828, τ=2.64 and (c) D*=1.379, τ=3. (B) We = 89. Inset: Front view post impact images depicting maximum wetting fraction for (a) D*=0.531, τ=3.31 (b) D*=0.828, τ=3.45 and (c) D*=1.379, τ=3.45. The magnitude of scale bar (inset c) is equal to 4 mm.

The second region of the wetting curve with respect to time is characterized by a change in slope, from the steep initial regime to a smoother segment. This region is dominated by viscous resistance to the inertial spreading, and the liquid mass approaches the south-pole with drastically reduced pace due to shear at the solid boundary. In the case of lower D*, the viscous shear is not potent enough to arrest the motion like high D*. Consequently, as the drop traverses towards the south pole, the gravity induced phase of spreading initiates, and a third regime of steep slope is observed in the lower D* cases just before the spreading maxima, or coalescence at the south pole, is observed. In addition to the wetting behaviour, the height of the liquid film at the north pole of the target is an important parameter which has been considered in the literature (Yarin & Wiess, 1995; Bakshi et al. 2007; Banitabaei & Amirfazli, 2017) for spherical targets. Fig 6 A and B illustrates the temporal behaviour of the non-dimensionalized film thickness at the north pole of the cylindrical target (described in fig 2b). Sharp drainage of the film is observed in the regime $0 < t^* < 0.8$, for all the cases, which conforms to earlier observations on curved surfaces (Banitabaei and Amirfazli 2017). The report also suggested that the behaviour can be described using $h^* = 1 - t^*$ for spherical targets and assuming that the velocity just after the impact remains unchanged. However, it has been observed that for cylindrical targets, this equation does not hold true and a new expression has been derived from first principles at a later stage in the article. After the initial phase of rapid depletion, there is no appreciable change in the film thickness. From fig. 6A it is observed that the film thickness after the initial phase remains higher for larger values of $D^*$. This trend is just opposite of what is noted for wetting fraction. Since for higher D*, the gravity induced film drainage phase near the south pole of the target is largely arrested, the effective thickness near the north pole remains higher. Additionally, for smaller D*, the lamellae also drains a considerable amount of fluid away during detachment, leading to low values of h*. As intuition suggests, the film thickness decreases for increasing impact We due to higher contribution due to inertia dominated spreading, which depletes the film at the north pole.

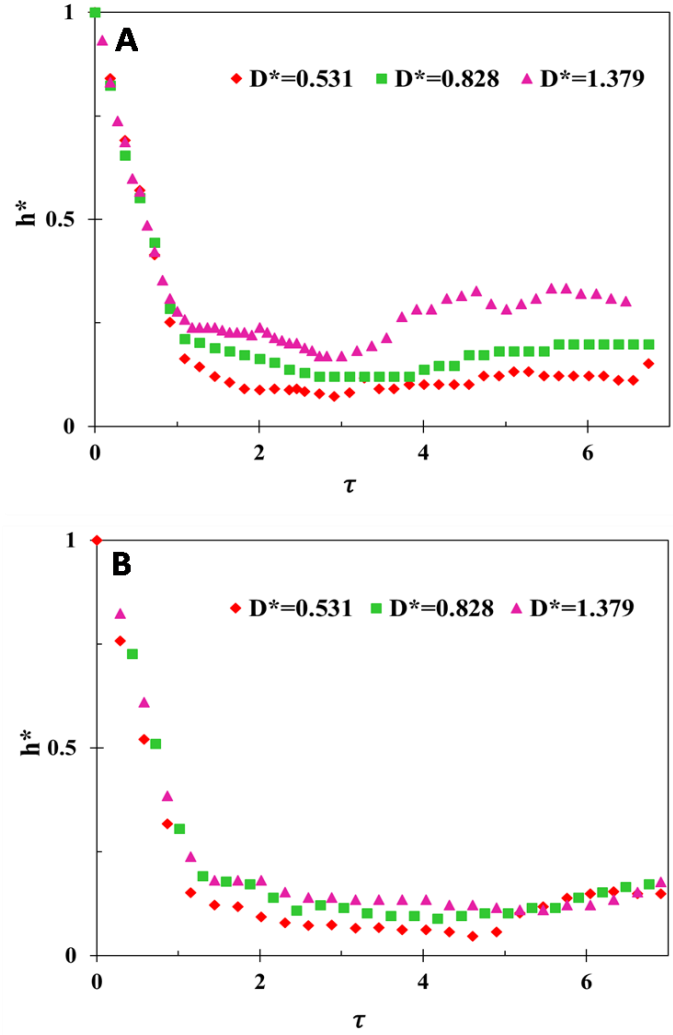

**Fig. 6:** Temporal variation of film thickness at the north pole of the hydrophilic cylindrical target surface of different D* (A) We=36. (B) We=89.

In fig. 6B, a plot similar to 6 A has been discussed, but for higher We (=89). It is observable that the effect of D* is largely diminished at higher impact velocities, for $t^* > 0.8$, whereas the trend in the initial deformation phase remains largely unchanged. The spreading factor (γ) is the final non-dimensional coefficient which has been employed to describe the post-impact hydrodynamics on curved surfaces. It is expressed as the ratio of post impact spreading length to initial droplet diameter (depicted in fig. 2 c). The role of D* on the spreading factor has been illustrated in Fig. 7 A and it is observed that the spreading factor increases with decreasing $D^*$. For curved systems, the spreading has been considered till the timeframe up to which the droplet does not distort to form the lamellae. The trends observed in fig. 7 A are consistent, and the spreading curves converge towards the flat surface case, which is typically $D^* = \infty$. Two distinct regimes are observed in the spreading evolution, divided by $\tau \sim \sqrt{8/3}$ (Sarojini et al. 2016), where the curves are observed to converge and then further diverge. This time instant represents the point at which an equivalent flat surface impact attains its maximum spreading diameter. The inset figures of fig. 7A represent spreading event for all the curved surfaces at that particular instant. It is interesting to note that at this instant, the spreading for all the surfaces have equal spreading factor of ~ 2.45 and the inset figures show that the shape of the droplet at this instant is fairly self-similar irrespective of the D*.

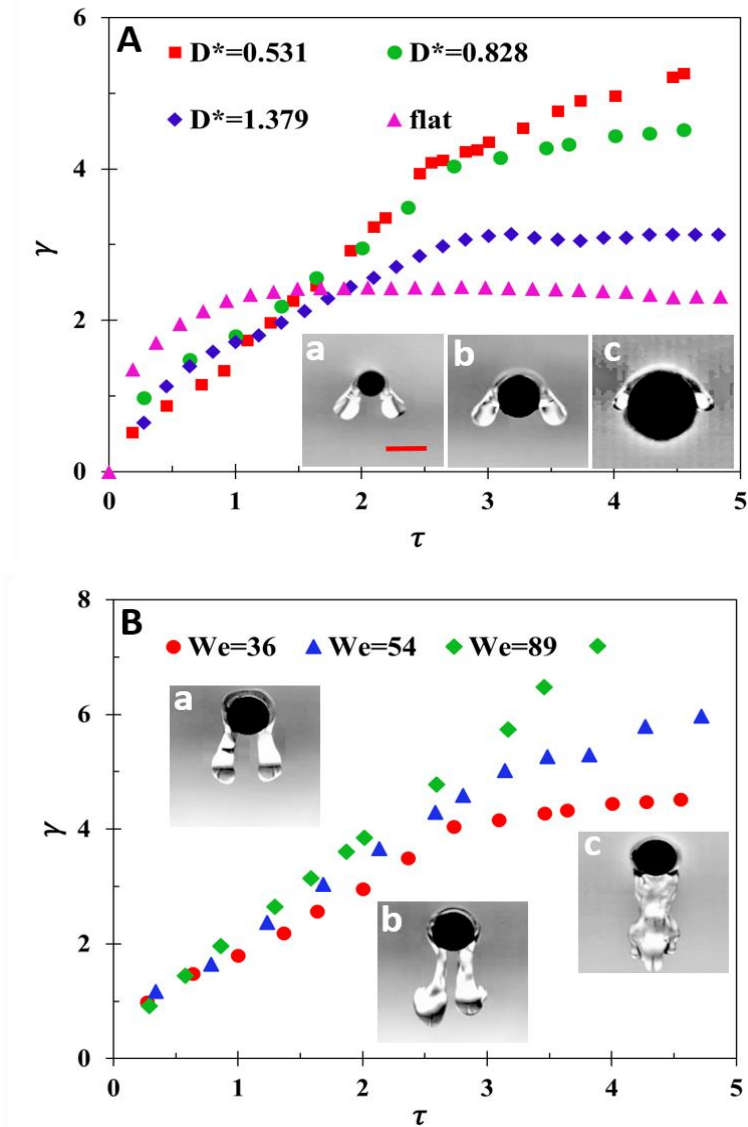

**Fig. 7:** (A) Temporal variation of spreading factor on hydrophilic cylindrical target surfaces of different D* for We= 36. Inset: Front view post impact images at $\tau$=1.63 for (a) D*=0.51

(b) D*=0.828 and (c) D*=1.379. (B) Temporal variation of spreading factor on hydrophilic cylindrical target surface (diameter 2.4mm) at different impact We. Inset: Front view post impact images at $\tau$=3.9 at impact We (a) 36 (b) 54 and (c) 89. The magnitude of scale bar is equal to 2.4 mm.

A distinct behaviour is observed in the time regime $\tau < \sqrt{8/3}$, where the spreading factor shows inverse proportionality to the value of D*, whereas in the regime $\tau > \sqrt{8/3}$, the opposite is noted. On a flat surface, the spreading event is only arrested by the viscous shear at the surface. In the case of curved surfaces, the initial inertial spreading is opposed not only by the viscous forces, but also by the curvature itself. For the same distance covered during spreading, the perimeter traversed is directly proportional to the diameter of the target. Accordingly, smaller D* values exhibit lower values of spreading in the initial time regime. Beyond this however, the spreading on flat system arrests due to balance of visco-capillary forces and the inertia. However, on a curved surface, the curvature promotes gravity induced spreading, which largely overcomes the viscous resistance. Since smaller perimeters provide lesser viscous resistance, the gravity induced spreading is much faster for lower D* targets. Fig. 7B illustrates the role of impact We on the spreading hydrodynamics for a constant D*. in this case, the curves show a diverging behaviour, with the spreading enhancing with increasing We. The figures in the inset depict spreading at the same time instant for all We cases illustrating the extent of spreading as function of We. It is interesting to note that up to the timescale $\tau \sim \sqrt{8/3}$, the spreading remains independent of the We. The role of the We initiates beyond this time, which signifies that the inertial regime of spreading is governed dominantly by the D* compared to the We.

### 3.2. Superhydrophobic surfaces

Similar experiments have been conducted on SH targets to understand the effect of wettability on the three geometrical parameters discussed. Fig. 8a shows a sequence of post impact images of droplet impact on a SH cylindrical target having 4 mm diameter and impact velocity of 1.5m/s. The impact outcomes are compared with fig 8b, which correspond to the equivalent hydrophilic target for the same impact conditions. While coalescence is observed in the hydrophilic case, for SH surfaces, the droplet is repelled off the surface in conjunction with severe deformation and fragmentation. This repulsive fragmentation and deformation is due to the interplay of the capillary forces (much higher $\sigma_{sl}$ of SH surfaces than hydrophilic surfaces) and the inertial forces. Comparative study of the arrays in 8a and 8b illustrates that the droplet on SH surface spreads along the surface, however largely without wetting the target, thereby forming elongated proto-lamellae (image 4 from left in array 8a and 8c). However, the lamellae do not develop like the hydrophilic case, but the droplet deforms and contorts while being repelled off the surface, often leading to fragmentation (last image from left in array 8a and 8c). Also, large brightness and contrast contours (for the same amount of backlight illumination as the hydrophilic cases) in the droplet in the SH case during proto-lamella formation and repulsive ejection can be observed. Such optical contrast is indicative of the large degree of surface capillary instabilities during the lamella formation and ejection away from the surface, which is comparatively less in the hydrophilic counterpart.

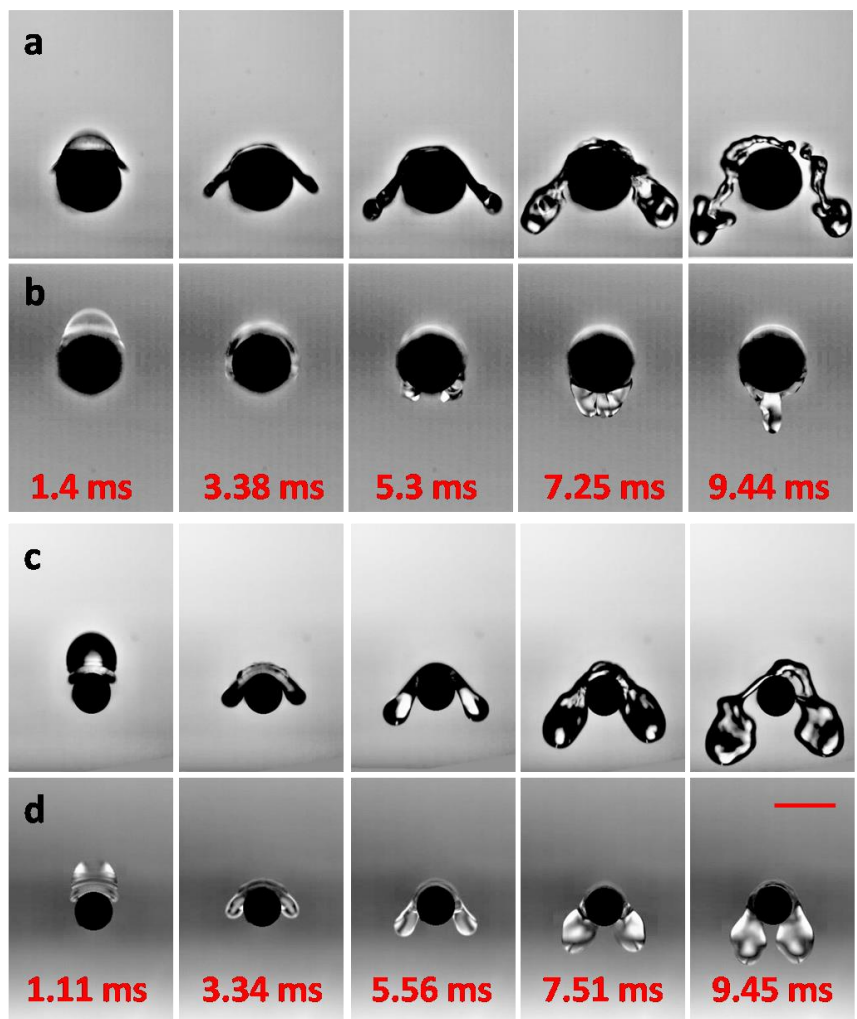

**Fig. 8:** Comparison of front view post impact images of droplet onto cylindrical target of different wettability having target diameter 4 mm at 1.5m/s impact velocity (We=89) (a) SH surface (b) Hydrophilic surface. Comparison of impact for target diameter 2.4 mm at 0.95 m/s impact velocity (We=36) (c) SH surface (d) Hydrophilic surface. The magnitude of scale bar is equal to 4 mm (bottom right).

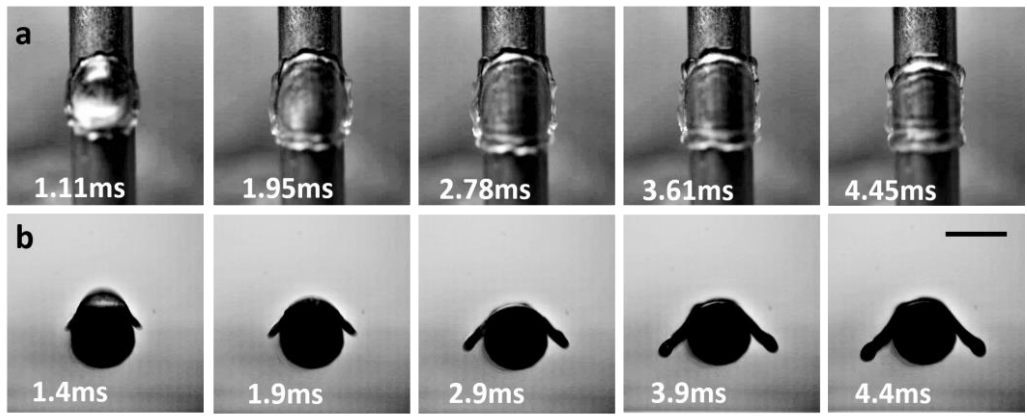

**Fig. 9:** Post impact images of onto SH cylindrical targets of diameter 4 mm at 1.5 m/s impact velocity (We=89) (a) top view (b) front view. The magnitude of scale bar is equal to 4 mm.

The largely augmented capillary instabilities of the deforming droplet on SH surfaces can be gauged from the simultaneous top view of the impact event. Fig. 9 shows the synchronized top and side views of the impingement process on SH cylindrical surface. From fig. 9, both the axial and azimuthal spreading can be observed. It can be seen that the spreading along the axial direction is arrested within the initial few microseconds (~2 ms) and the spread of the drop along the cylinder axis is absent thereafter. The top view illustrates that a rimmed edge is formed around the droplet during the spreading event, and capillary instability or capillary waves are visible during the film drainage phase. When compared to the impact events on a hydrophilic target, the wetting fraction is found to be significantly reduced for SH targets (fig. 10 A) due to the high solid-liquid surface energy component of SH surfaces. On the contrary, the spreading factor and non-dimensional film thickness remain very similar for both wettability targets (fig 10 B & C). However, the phenomena of spreading without wetting creates significant differences compared to spreading with wetting, which can be further seen in the arrays in fig. 8. Images shown in fig 10 D and E compare the top view images of droplet impact on hydrophilic and SH surfaces, respectively, under the same impact conditions. Interestingly, the top views have very similar structures and the axial spread is also very similar. Such observations signify that the major differences in hydrodynamics due to wettability is prominent only as the focus shifts towards the south pole of the target, and at the north pole and its vicinity, impact creates hydrodynamic events which are independent of wettability.

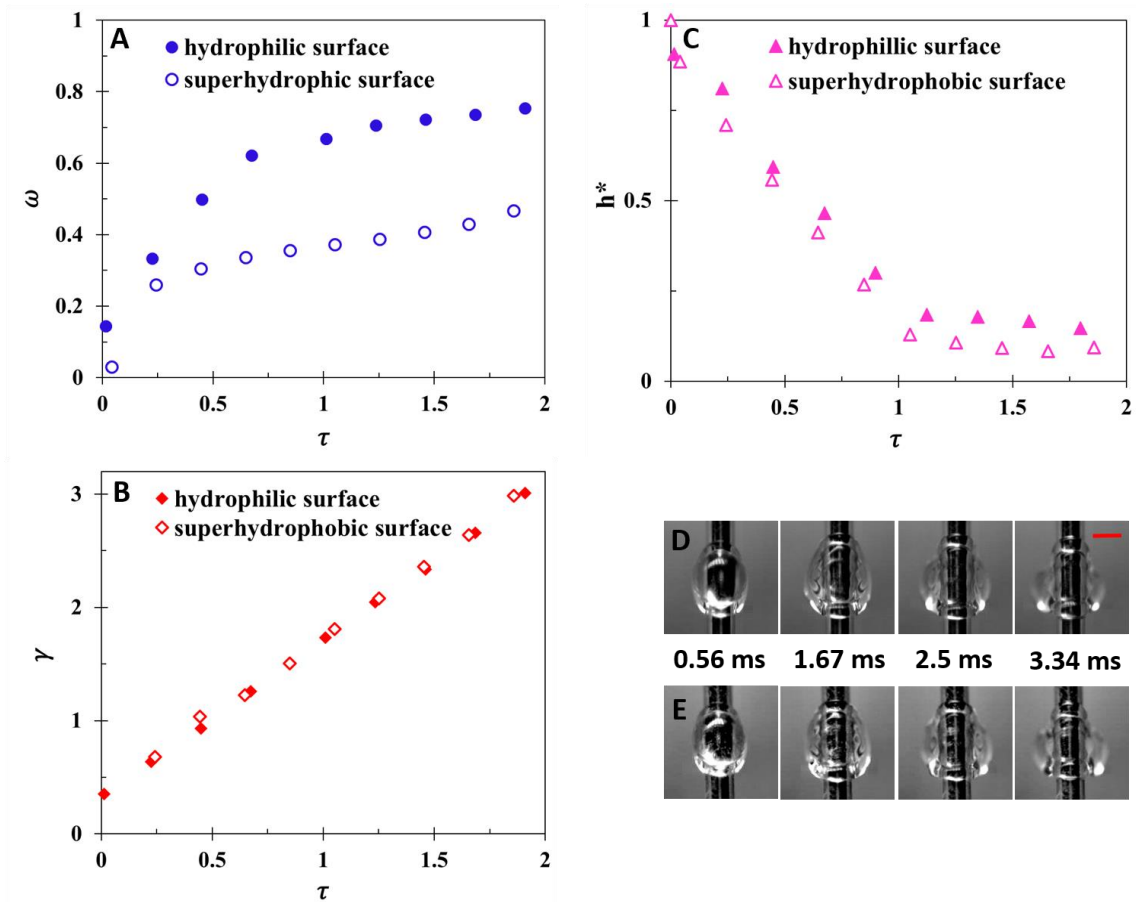

**Fig. 10:** Temporal variations of (a) wetting fraction (b) spreading factor (c) non-dimensional film thickness at the north pole of the cylindrical target for different wettability. Comparison of top view post impact images for target diameter 1.54 mm and 1.17 m/s impact velocity (We=54) for (d) hydrophilic surface (e) SH surface. The magnitude of scale bar is equal to 1.54mm.

### 3.3. Mathematical formulation

The present section discusses a mathematical formulation which has been deduced from first principles to model the hydrodynamics of the droplets post impact. The evolution of the non-dimensional film thickness has been derived based on the analytical domain illustrated in fig. 11. The film evolution process is considered to happen in accordance to the geometry specified in fig. 11, with minimal flow along the direction of the cylinder axis. This phenomenon has been observed from top view analysis and it is observed that the lateral spread of the droplet during film evolution is negligibly small.

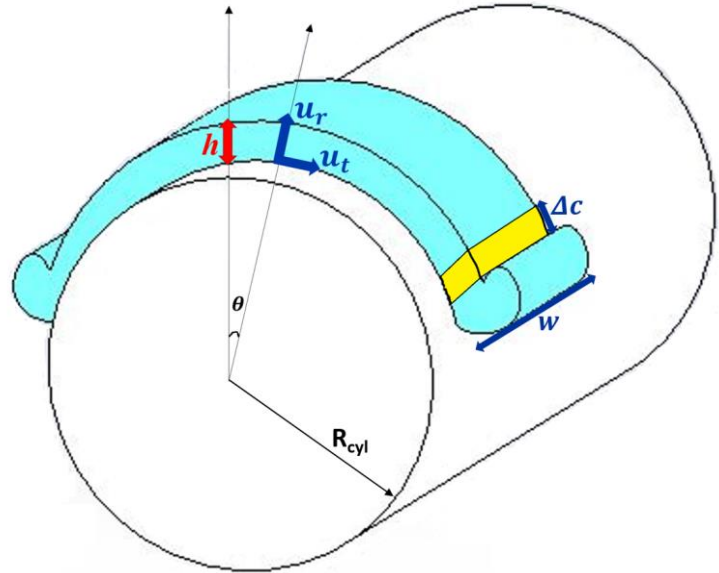

**Fig. 11:** Analytical description of the film flow on the cylindrical target during evolution of the film. The similarity of the simplified geometry considered can be observed from Fig. 3, We = 36, 3.06 ms and Fig. 10 (d) 2.5 ms.

During the film spreading regime, $(0.4 < t^* < 2.9)$, the role of viscosity is considered to be minimal and no hydrodynamic stress exists along the tangential direction, and the droplet is assumed to spread in the inertial regime. The spread of the elemental region, c, with time can be expressed as

$$\frac{\partial c}{\partial t} = u_t \qquad (1)$$

$$\frac{\partial u_t}{\partial t} = 0 \qquad (2)$$

where $c = R_{cyl}\theta$ and $u$ represents the velocity and subscripts $t$ and $r$ representing tangential and radial, respectively. Additonally, as expressed in eqn. 2, it is further assumed that the film of fluid spreads uniformly over the cylinder without any appreciable acceleration or decelerative components. The equations are subjected to the initial conditions $c = c_0$ and $u_t = u_0(c_0)$ at $t = t_i$, where $t_i$ is the time instant after which the flow can be considered as a film flow and the negligible viscous resistance can be employed. The generic solutions can be expressed as

$$c = c_0 + u_0(c_0)(t - t_i) \text{ and } u_t = u_0(c_0). \tag{3}$$

Further, considering a film element of infinitesimal length $\Delta c$ as shown in figure 11, the analytical form of the system can be realized. The volume of this film element is $\Delta V = wh\Delta c$, where $w$ and $h$ represent the width and thickness of the fluid element, respectively. The analysis assuming $w$ to be constant throughout the film length, which is also observed to be true experimentally during the film spreading regime. By applying conservation of mass to this film element, the evolution of the film can be expressed as

$$h\Delta c = h_0(c_0)\Delta c_0 \tag{4}$$

where $h_0(c_0)$ is film thickness at instant $t = t_i$. At the limit of vanishing element size, $\Delta c_0 \rightarrow 0$, the generic expression in eqn. 4 can be expressed as

$$h(c_0, t) = h_0(c_0)\frac{\partial c_0}{\partial c}. \tag{5}$$

From the typical solution expressed in eqn. 3, the eqn. 5 can be further written in the form as

$$h(c_0, t) = \frac{h_0(c_0)}{1 + \frac{du_0}{dc_0}(t - t_i)} \tag{6}$$

The solution for the differential equation maybe achieved employing the concept of remote asymptotic solution (Bakshi et al. (2007)). From the geometry, the tangential velocity immediately after impact conditions can be expressed as

$$u_t = U\sin\theta \tag{7}$$

where $U$ is the droplet impact velocity. At the vertical axis $(c_0 = 0)$ $u_t = 0$ therefore, the initial velocity and initial film thickness can be approximated as a linear function of $c_0$, expressible as

$$u_0 \approx A c_0 \text{ and } h_0(c_0) \approx h_0(c_0 = 0) \tag{8}$$

where $A$ is a constant. Upon combining equations (3), (5) and (8), the expressions for the temporally evolving film thickness and the tangential velocity is obtained as

$$h(t) = \frac{h_0}{1 + A(t - t_i)} \tag{9}$$

$$u_t = \frac{A R_{cyl} \theta}{1 + A(t - t_i)} \tag{10}$$

In order to obtain the values of determine the parameters $h_0$, $A$ and $t_i$, further analysis of the geometrical evolution is necessary. For regions very near the vertical axis $(c_0 = 0)$, $u_t \approx U\theta$. Comparing equation (8) with this near-axis observation, it can be shown that

$$A = \frac{U}{R_{cyl}} \tag{11}$$

At $t = t_i$, $h$ is substituted from equation (9) for the invisid film distribution stage, and a similar relationship can be obtained as

$$\frac{dh}{dt} = -U, \tag{12}$$

Also, the initial value of h (just before the film spreading regime initiates) can be expressed as

$$h_0 = R_{cyl} \tag{13}$$

at $t = t_i$ for initial impact stage. Hence the solution of equation 12 is expressed as

$$h = D - Ut \tag{14}$$

Where D is the diameter of the impacting droplet. Employing the smooth matching condition for film thickness at $t = t_i$, the time can be further deduced as

$$t_i = \frac{D - R_{cyl}}{U} \tag{15}$$

Upon substituting the expressions for the obtained parameters in equation (9), the final expression is obtained as

$$h(t) = \frac{R_{cyl}^2}{2R_{cyl} + Ut - D} \tag{16}$$

The expression for the non-dimensionalized film thickness is as

$$h^*(t) = \frac{R_{cyl}^2}{2R_{cyl}D + D^2 t^* - D^2} \tag{17}$$

The theoretically deduced values of the film thickness have been compared against the experimentally observed values, for different We, D* and wettability values, in Fig. 12, and good agreement has been observed between the two data sets.

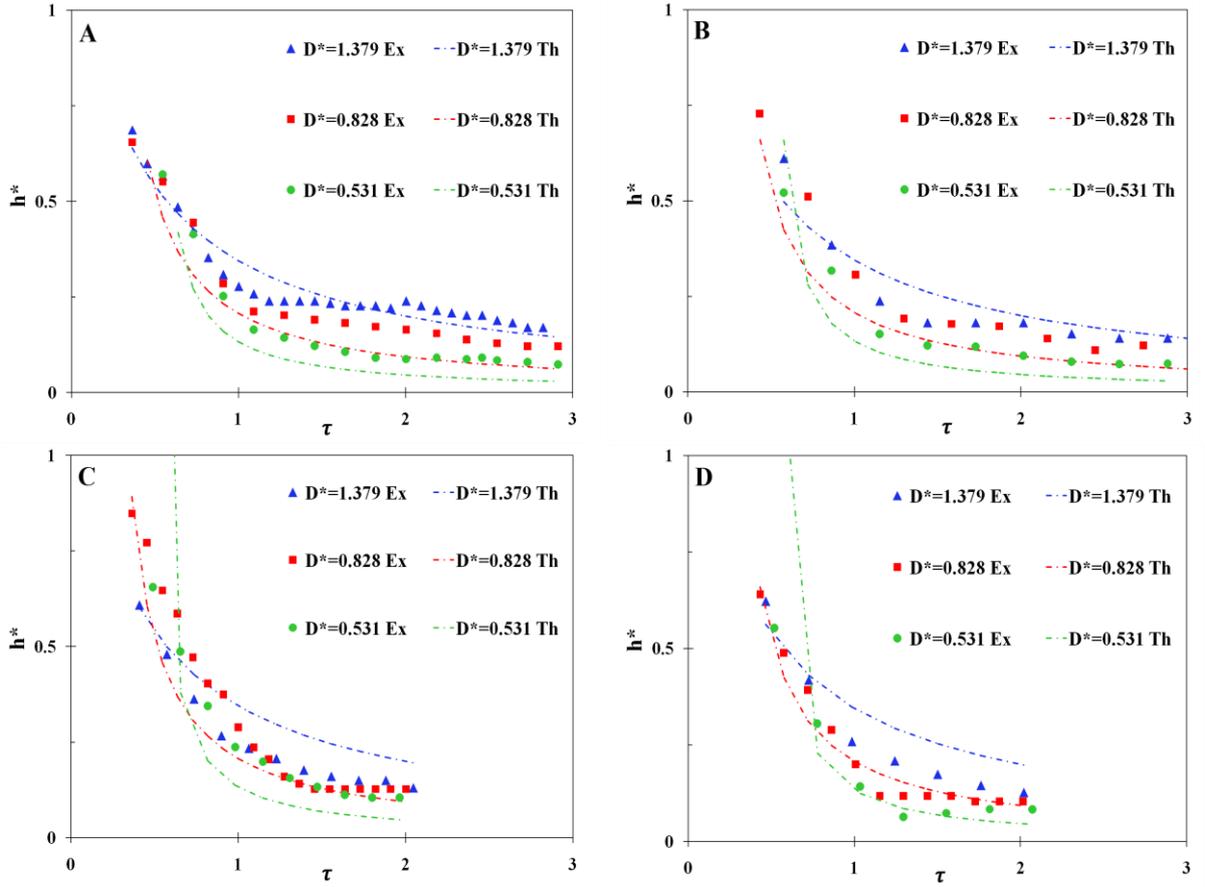

**Fig. 12:** Comparision of experimental and theoretical film thickness evolution of water droplets at the north pole for (a) We=36 on hydrophilic surface, (b) We= 89 on hydrophilic surface, (c) We=36 on SH surface and (d) We= 89 on SH surface.

In addition to modelling the film thickness evolution during the spreading regime, a theoretical model to determine the maximum wetted angle ($2\theta$) on the cylindrical body has also been proposed. Based on energy balance principle, the pre and post impact energies are considered to be conserved, and expressed as

$$k.e_i + s.e_i = s.e_1 + s.e_2 + s.e_3 + s.e_4 + v_{dis} \tag{18}$$

Where, the kinetic energy pre-impact can be expressed as

$$k.e_i = \frac{1}{2}\left(\rho U^2\right)\left(\frac{\pi d^3}{6}\right) \tag{19}$$

Considering that the droplet before impact assumes a perfectly spherical shape (experiments reveal that the droplets are nearly spherical before impact, however, mild distortions are possible, which are neglected in the present case), the surface energy of the droplet pre-impact is expressed as

$$s.e_i = \pi d^2 \sigma_{lg} \tag{20}$$

Post-impact, the energies are a sum of the variant surface energy components and the viscous dissipation work done by the droplet while spreading against the surface. The viscous dissipation work can be expressed as

$$v_{dis} = \int_0^{t_c} \int_V \psi \, dV \, dt \tag{21}$$

The dissipative energy can be expressed as a function of the impact velocity and the analogous boundary layer thickness δ as

$$\psi \approx \mu \left(\frac{U}{\delta}\right)^2 \tag{22}$$

Based on a boundary layer flow analogy over a flat plate (it is assumed that the spreading process is largely inertia driven such that the curvature effects do not give rise to pressure gradients), the boundary layer thickness can be scaled as

$$\delta \simeq \frac{5 R_{cyl}(2\theta)}{\sqrt{Re}} \tag{23}$$

where $Re = \rho U R_{cyl}(2\theta)/\mu$. ~

The effective volume of the liquid under the action of the viscous dissipative work can be expressed as

$$V_{dis} = R_{cyl}(2\theta) \, w \, \delta \tag{24}$$

Where the width of the spreading liquid layer is of the form

$$w = \frac{\pi}{6} \frac{d^3}{R_{cyl}(2\theta) \mathrm{h}_{\theta/2}} \tag{25}$$

In Eqn. 25, $\mathrm{h}_{\theta/2}$ denotes the liquid film thickness at the time instant of maximum spreading at polar location $\theta/2$. The average time of spreading, during which the viscous forces are dominant can be scaled as

$$t_{spread} \sim \frac{R_{cyl}\theta}{U} \tag{26}$$

Integrating eqn. 21 with the expressions obtained from eqns. 22–26, the final form of the viscous dissipation work component can be expressed as

$$v_{dis} = \frac{\pi d^3}{6} \frac{U \mu R_{cyl}(2\theta)}{2 h_{\theta/2} \delta} \tag{27}$$

In addition to the viscous work, the components of surface and interfacial energies also play a governing role on the droplet hydrodynamics post impact. The component of surface energy due to the solid-liquid interface during the film drainage regime can be expressed as

$$s.e_1 = \frac{\pi d^3 \sigma_{sl}}{6 h_{\theta/2}} \tag{28}$$

Likewise, the component of surface energy due to the liquid-gas interface during the film drainage regime is of the form

$$s.e_2 = \frac{\pi d^3 \sigma_{lg}}{6 h_{\theta/2}} \tag{29}$$

In the spreading regime, the component of surface energy due to the liquid-gas interface is of the form

$$s.e_3 = \frac{\pi d^3 \sigma_{lg}}{3 R_{cyl}(2\theta)} \tag{30}$$

While the surface energy in the spreading regime due to the solid-liquid interface is expressed as

$$s.e_4 = 2 R_{cyl}(2\theta) h_{\theta/2} \sigma_{lg} \tag{31}$$

Substituting all the energy components in eqn. 18, the net energy balance equation is of the form

$$\frac{1}{2}\left(\rho U^{2}\right)\left(\frac{\pi d^{3}}{6}\right)+\pi d^{2}\sigma_{lg} = \frac{\pi d^{3}}{6h_{\theta/2}}\left(\sigma_{sl}+\sigma_{lg}\right)+\frac{\pi d^{3}\sigma_{lg}}{3R_{cyl}(2\theta)}+2R_{cyl}(2\theta)h_{\theta/2}\sigma_{lg}+\frac{\pi d^{3}}{6}\frac{U\mu R_{cyl}(2\theta)}{2h_{\theta/2}\delta} \quad (32)$$

Non-dimensionalizing the expression with respect to the initial surface energy of the pre-impact droplet, the relationship between the spreading angle is obtained in terms of the governing Weber (We) and Capillary (Ca) numbers as

$$\frac{We}{12}+1 = \frac{Ca}{12}\frac{R_{cyl}d(2\theta)}{h_{\theta/2}\delta}+\frac{d}{6h_{\theta/2}}\left(1+\frac{\sigma_{sl}}{\sigma_{lg}}\right)+\frac{d}{3R_{cyl}(2\theta)}+\frac{2R_{cyl}(2\theta)h_{\theta/2}}{\pi d^{2}} \quad (33)$$

Which can be further expressed as a quadratic function of $2\theta$ as

$$\left(\frac{CaR_{cyl}d}{12h_{\theta/2}\delta}+\frac{2R_{cyl}h_{\theta/2}}{\pi d^{2}}\right)2\theta^{2}+\left(\frac{d}{6h_{\theta/2}}\left(1+\frac{\sigma_{sl}}{\sigma_{lg}}\right)-\frac{We}{12}-1\right)2\theta+\frac{d}{3R_{cyl}}=0 \quad (34)$$

The equation 34 is solved and the theoretical values are compared with the experimentally observed values of $2\theta$ and the percentage errors between the predictions have been illustrated in fig 13 for representative impact and wettability conditions. It is observed that for hydrophilic surfaces, the model performs better in case of higher D* and lower We regimes, whereas for SH surfaces, the performance is better for higher D* as well as higher We regimes. In case of lower D*, the droplet undergoes rapid film drainage caused by the large curvature, which reduces the components of surface energies due to film drainage, thereby reducing the predictability of the model. On hydrophilic bodies, the high We regime leads to enhanced splashing component of surface energy, which is not accounted for in the model, and hence the deviations enhance for high We on hydrophilic surface. On the contrary, on SH surfaces, higher inertia ensures proper spread of the droplet before the capillary forces breaks the droplet up into daughter droplets. At lower We, the spread of the droplet is insufficient due to lower inertia, and hence the model performs poorly for low We regime on SH surfaces.

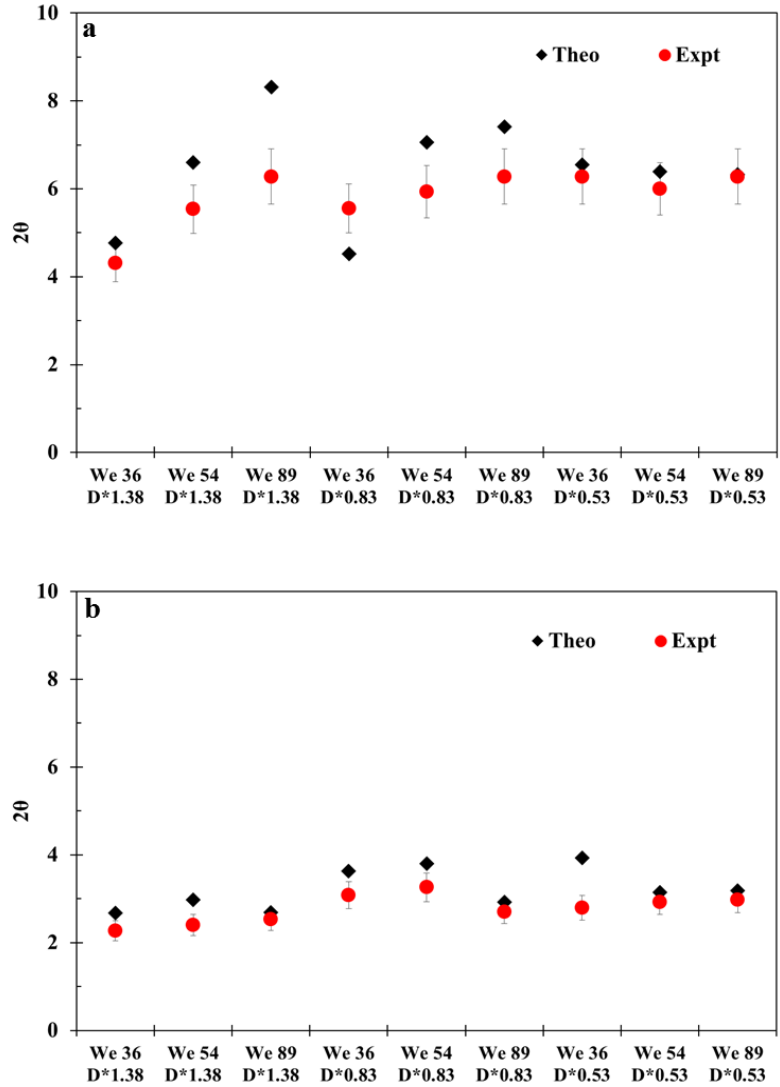

**Fig. 13:** Comparison of experimental and theoretical values of $2\theta$ for various impact conditions on cylindrical targets with (a) hydrophilic surface and (b) SH surface

## 4. Conclusions

The present article experimentally investigates the impact dynamics and post-impact hydrodynamics of water droplets on cylindrical targets of different diameters and wettability. The study involves variant impact We and hydrodynamic behavior such as wetting, spreading, film drainage and lamellae formation and behavior have been studied in depth using high speed photography, both from side and top views. It is observed that when the target diameter is increased with respect to the droplet size at the same impact We, the wetting fraction and spread factor decrease, whereas the film thickness increases. Upon increasing the impact We, the effect of curvature on wetting fraction, spread factor and film thickness is observed to reduce. It is also observed that the maximum wetting fraction is attained faster for lower diameter ratios. Moreover, in the case of SH surfaces, the wetting fraction reduces significantly, whereas the spread factor remains comparable to that of hydrophilic surfaces. It is also shown that the diameter ratio and impact We has a direct influence on the lamellae shape, growth behavior, formation and break-off dynamics, as well as on its instability behavior during breakup mode. Additionally, an analytical expression for temporal evolution of the film thickness at the north pole of the target has been presented and good agreement is achieved with respect to experimental data. Having discussed the morphology of post impact droplet, a theoretical model based on energy conservation principle has also been proposed to predict the maximum wetting fraction on a given cylindrical target surface. The model predicts the spreading of the droplet over the curved surface in terms of the governing We and Ca. The model is observed to be able to predict the experimental results with appreciable accuracy, considering the highly dynamic and statistical nature of such impact and wetting phenomena. The present study thus sheds good insight on the hydrodynamics of droplets and post-impact events on cylindrical targets and illustrates the effects of diameter ratio, impact We and Ca and wettability on the post impact fluid dynamics.

## Acknowledgements

GK and NS thank the Ministry of Human Resource Development, govt. of India, for the doctoral fellowships. PD thanks IIT Ropar for funding the present research (vide grants IITRPR/Interdisciplinary/CDT and IITRPR/Research/193).